\documentstyle[prl,aps,multicol,psfig,epsfig,pstricks]{revtex}  
\begin{document}
\newcommand{\be}{\begin{equation}}
\newcommand{\ee}{\end{equation}}
\newcommand{\bq}{\begin{eqnarray}}
\newcommand{\eq}{\end{eqnarray}}
\newcommand{\fat}[1]{\mbox{\boldmath $ #1 $\unboldmath}}
\newcommand{\puteps}[5]{
\begin{center}
\parbox[t]{#2}{
\begin{figure}[hbt]
{\psfig{file= #1 ,angle=0,width= #2 ,height= #3}}
\caption{ #4 }
\label{#5}
\end{figure}}
\end{center}}

\title{Scaling of Entanglement close to a Quantum Phase Transitions}
\author{A. Osterloh$^{(1)}$, L. Amico$^{(1)}$, G. Falci$^{(1)}$, 
\& R. Fazio$^{(2)}$}
\address{(1)\, NEST-INFM $\&$ Dipartimento di Metodologie Fisiche e Chimiche (DMFCI),
 viale A. Doria 6, 95125 Catania, ITALY}
\address{(2)\, NEST-INFM $\&$ Scuola Normale Superiore, I-56127 Pisa, ITALY}

\maketitle

\begin{abstract}
In this Letter we discuss the entanglement near a quantum phase 
transition by analyzing the properties of the concurrence for a class of 
exactly solvable models in one dimension. We find that entanglement 
can be classified in the framework of scaling theory.
Further, we reveal a profound difference between classical correlations 
and the non-local quantum correlation, entanglement: the
correlation length diverges at the phase transition, whereas 
entanglement in general remains short ranged.
\end{abstract}
%\PACS 02.30.Ik , 74.20.Fg , 03.65.Fd


\begin{multicols}{2}

\narrowtext
Classical phase transitions occur when a physical system reaches a state 
below a critical temperature characterized by a macroscopic 
order~\cite{PHASE-TRANS}. 
Quantum phase transitions occur 
at absolute zero; they are induced by the change of an 
external parameter or coupling constant~\cite{QPHASE-TRANS}, and are 
driven by fluctuations. 
Examples include transitions
in quantum Hall systems~\cite{QHALL}, localization in 
Si-MOSFETs (metal oxide silicon field-effect transistors; Ref.~\cite{M-I}) 
and the superconductor-insulator 
transition in two-dimensional systems ~\cite{2D,S-I}. 
Both classical and quantum critical points are governed by a 
diverging correlation length, although quantum systems possess 
additional correlations that do not have a classical counterpart. 
This phenomenon, known as entanglement~\cite{FOUNDATION}, 
is the resource that enables quantum computation and communication~\cite{QCOMPUTATION}. 
The role of the entanglement at a phase transition is not captured
by statistical mechanics -- 
a complete classification of the critical many-body state
 requires the introduction of concepts from quantum information 
theory~\cite{PRESKILL}.
Here we connect the theory of critical phenomena 
with quantum information by exploring the entangling resources of a 
system close to its quantum critical point.
We demonstrate, for a class of one-dimensional magnetic systems,
that entanglement shows scaling behaviour in the vicinity of 
the transition point.

There are various questions that emerge in the study of this problem. 
Since the ground state wave-function undergoes qualitative changes 
at a quantum phase transition, it is important to understand how 
its genuine quantum aspects evolve throughout the transition. 
Will entanglement between distant subsystems be extended
 over macroscopic regions, as correlations are? 
Will it carry distinct features of the transition itself  and 
show scaling behaviour? Answering these questions is important for a 
deeper understanding of quantum phase transitions and also from the 
perspective of quantum information theory.
So results that bridge these two areas
 of research are of great relevance.
We study a set  of localized spins coupled through exchange interaction 
and subject to an external magnetic field (we consider only spin-1/2 
particles), a model central both to condensed matter and 
information theory and subject to intense study~\cite{BROOKE}. 
\puteps{fig1.eps}{8cm}{5cm}{
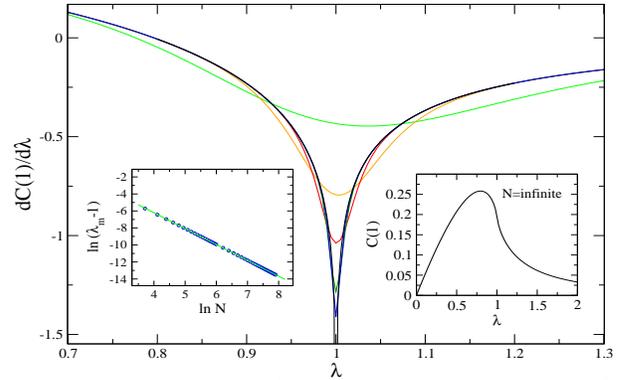
The change in the ground state wave-function 
in the critical region 
is analyzed considering $\partial_\lambda C(1)$ 
as a function of the reduced coupling 
strength $\lambda$. The  curves correspond to different lattice sizes
$N=11,41,101,251,401,\infty$. We choose $N$ odd to avoid the 
subtleties connected with boundary terms\protect\cite{DMRG}. 
On increasing the system size,
the minimum gets more pronounced. also the position of the minimum 
changes and tends as $N^{-1.87}$ (left inset) towards the critical 
point $\lambda_c =1$ where for an infinite system a logarithmic divergence
is present (see equation (\protect\ref{nn-concurrence})). 
The right inset shows the behaviour 
of the concurrence $C(1)$ itself for an infinite system. The maximum that 
occurs below $\lambda_c$ is not related to the critical properties
of the Ising model. As explained in the text, it is the 
change in the ground state and not the wavefunction itself
that is a good indicator of the transition. 
The structure of the reduced density matrix, necessary to calculate 
the concurrence follows from the symmetry properties of the Hamiltonian. 
Reality and parity conservation of $H$ together with translational invariance 
fix the structure of  $\rho$ to be
real symmetric with $\rho_{11}$, 
$\rho_{22}=\rho_{33}$, $\rho_{23}$, $\rho_{14}$, $\rho_{44}$ as the only 
non-zero entries.}{fig1}

O'Connors and Wootters showed that in the Heisenberg 
chain the maximization of the entanglement at zero temperature 
is related to the energy minimization \cite{RINGS}. It is known that 
Werner states~\cite{3-QBITS} can be generated in a one dimensional 
XY model~\cite{HEISENBERG} and that temperature 
and magnetic field can increase the 
entanglement of the systems as shown for the Ising and Heisenberg 
models~\cite{THERMAL,MIXING}. Finally we mention the study on the role of 
the entanglement in the density matrix renormalization group 
flow and the introduction of entanglement-preserving renormalization 
schemes~\cite{DMRG}. Here we address the problem of the relation 
between macroscopic order, classical correlations, and quantum correlations. 
Therefore we analyze the entanglement near the critical point of the XY 
model in a transverse field. 
Because of the universality principle -- 
 the critical behaviour depends only on the dimension of the
 system and the symmetry of the order parameter -- our results have much 
broader validity. 
%The concurrence, 
%a measure for the resource entanglement~\cite{CONCURRENCE}, 
%is calculated exactly for generic number of spins. 
We find that in the vicinity of a quantum phase 
transition the entanglement obeys scaling behaviour. 
On the other hand this analysis provides a 
clear distinction between the role of entanglement and correlations 
in quantum systems close to a critical point.
(We have been made aware that similar work to that reported here is 
being performed by T. Osborne and M. Nielsen; T. Osborne, 
personal communication.\cite{OSBORNE}) 
\puteps{fig2.eps}{8cm}{5cm}{
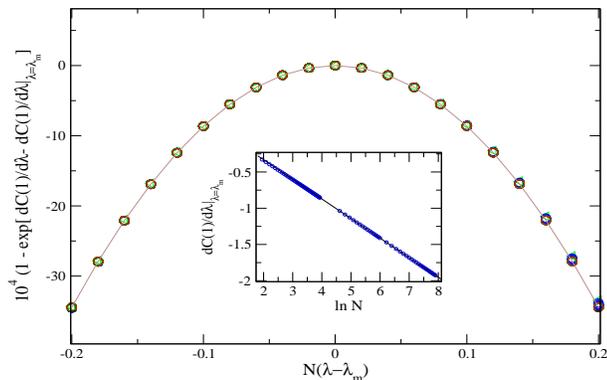
The finite size scaling is performed for the case of 
logarithmic divergences\protect\cite{LIEB}. 
The concurrence, considered as a function of the system 
size and the coupling constant, is a function of
$N^{1/\nu}(\lambda-\lambda_m)$ only, and 
in the case of log divergence it behaves as  
\(
\partial_\lambda C(1)(N,\lambda)\) --
\(\partial_\lambda C(1)(N,\lambda_0)  \sim\)
\(Q[N^{1/\nu}(\lambda-\lambda_m)]\) -- \(Q[N^{1/\nu}(\lambda_0-\lambda_m)]
\)
where $\lambda_0$ is a non-critical value and 
$Q(x)\sim Q(\infty) \ln x$ (for large $x$). 
All the data  from $N=41$ up to $N=2701$ collapse on a single curve.
The critical exponent is $\nu=1$, as expected for the Ising model. 
The inset shows the divergence of the value at the minimum as the 
system size increases.}{fig2}

The system under consideration is a spin-1/2 ferromagnetic chain with
 an exchange coupling J in a transverse magnetic field of strength $h$. 
The Hamiltonian is 
\begin{equation}
H= \mbox{--} \frac{J}{2}\sum_{i=1}^N (1 \mbox{--} \gamma)\sigma^x_i \sigma^x_{i+1}+
(1+\gamma)\sigma^y_i \sigma^y_{i+1} \mbox{--} h\sum_{i=1}^N \sigma^z_i
\label{model} 
\end{equation}
where $\sigma^a$ are the Pauli matrices ($a=x,y,z$) and $N$ is the number of
 sites. We assume periodic boundary conditions. 
It is convenient, for later purposes, to define a dimension-less 
coupling constant $\lambda=J/2h$. For $\gamma =1$ Eq.~(\ref{model}) 
reduces to the Ising model whereas for $\gamma = 0$ it is the XY model. 
For all the interval  $0<\gamma\le 1$ the models belong to the Ising 
universality class and for $ N =\infty$ they undergo a quantum phase 
transition at the critical value $\lambda_c=1$. The magnetization $\langle 
\sigma^x\rangle $ is different from zero for $\lambda >1$ and it vanishes 
at the transition. On the contrary the magnetization along the 
$z$-direction 
$\langle \sigma^z\rangle $ is different from zero for any value of $\lambda$. 
At the phase transition the correlation length 
$\xi$ diverges as $\xi \sim |\lambda - \lambda_c |^\nu$ with 
$\nu=1$ (Refs. \cite{PFEUTY} and \cite{McCOY}).
We confine our interest to the entanglement between two spins,
on position $i$ and $j$, in the chain. All the information needed is 
contained in the reduced density matrix $\rho(i,j)$ obtained from the 
ground state wave-function after all the spins except those at positions $i$ 
and $j$ have been traced out. 
The resulting $\rho(i,j)$ represents a mixed state of a bipartite system;
a good deal of work has been devoted to quantify the entanglement
in this case\cite{CONCURRENCE,ENTANGL}. We use the concurrence
between sites $i$ and $j$,  
related to the ``entanglement of formation'', and defined as~\cite{CONCURRENCE}  
\begin{equation}
C(i,j)=\max\{0,r_1(i,j)\mbox{--}r_2(i,j)\mbox{--}r_3(i,j)\mbox{--}r_4(i,j)\}
\label{concurrence}
\end{equation} 
In Eq.(\ref{concurrence}) $r_\alpha(i,j) $ are the square roots of 
the eigenvalues of the product matrix $R=\rho (i,j) \tilde{\rho} (i,j) $ 
in descending order; 
the spin flipped matrix $\tilde{\rho}$ is defined as 
$\tilde{\rho}\doteq \sigma^y\otimes\sigma^y \rho^* \sigma^y\otimes\sigma^y$. 
In the previous definition, the eigenstates of $\sigma^z\in 
\{|\uparrow \rangle \, |\downarrow \rangle\}$ should be used. 
Translation invariance implies that $C(i,j)=C(|i-j|)$. 
The concurrence will be evaluated as a function of the relative 
position $|i-j|$ between the spins and the distance 
$|\lambda -\lambda_c|$ from the critical point. 
The structure of the reduced density matrix is obtained 
by exploiting symmetries of the model (see caption of Fig. 1). 
The non-zero entries of $\rho$ can then be related to the various 
correlation functions, and the concurrence of the ground state is 
evaluated exactly starting from the results in Refs.~\cite{PFEUTY},
\cite{McCOY}, and \cite{LIEB}.

First we look at the Ising Model ($\gamma=1$).
The first question we consider is the range of 
the entanglement $\xi_E$, that is, the maximum distance 
between two spins at which the concurrence is different from zero. 
The result is surprising: even at the critical point, where spin-spin 
correlations extend over a long range (the correlation length is 
diverging for an infinite system), 
the concurrence vanishes unless the two sites are at most 
next-nearest neighbors. The truly non-local quantum part of the 
two-point correlations is nonetheless very short-ranged.
 
In order to quantify the change of the many-body wave-function when 
the system crosses the critical point, we look at the derivatives of
the concurrence as a function of $\lambda$. In this case we need to
consider only the nearest neighbour and next nearest neighbour
concurrence. We first discuss the behaviour of the nearest 
neighbor concurrence. 
\puteps{fig3.eps}{8cm}{5cm}{
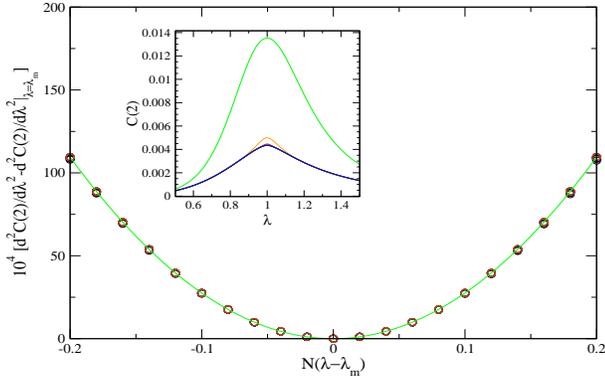
As in the case of the nearest neighbour concurrence, data 
collapse is also obtained for the next-nearest-neighbour concurrence C(2).
In the figure, data for system size from $N= 41$ to $N=401$ are plotted. The 
inset shows a peculiarity of the Ising model: 
$C(2)$ has its maximum precisely at the 
critical point for arbitrary system size (note that the maximum decreases
as the system size increases). 
Therefore we consider the second derivative to perform the scaling analysis.
It can also be seen that $C(2)$ is two orders of magnitude smaller than $C(1)$.
For the smallest system sizes the concurrence
is different from zero for $|i-j| = 3$ and $\lambda > 1.05$ (for $N=7$; 
for $N\ge 9$, $C(3)=0$ for all $\lambda$). 
In contrast the correlation functions are long-ranged at the critical point.}
{fig3}
The results for systems of different 
size (including the thermodynamic limit) are presented in Fig.\ref{fig1}.
For the infinite chain $\partial_\lambda C(1) $ 
diverges on approaching the critical value as 
\begin{equation}
\partial_\lambda C(1)=\frac{8}{3 \pi^2} \ln|\lambda-\lambda_c| +const.
\label{nn-concurrence}
\end{equation}      				
Equation (\ref{nn-concurrence}) quantifies non-local correlations 
in the critical region. 
One aspect of this system, particularly relevant 
for quantum information is the study of the precursors of the 
critical behaviour in finite samples. This study is known
as finite size scaling~\cite{FINITE-SCALING}. 
In Fig.\ref{fig1} the derivative of $C(1)$ 
respect to $\lambda$ is considered for different system sizes. 
As expected, there is no divergence for finite $N$, but there are clear 
anomalies. The position of the minimum $\lambda_m$ scales as 
$\lambda_m \sim \lambda_c + N^{-1.87}$ and its value diverges logarithmically 
with increasing system size as 
\begin{equation}
\left. \partial_\lambda C(1)\right |_{\lambda_m}=-0.2702 \ln N +const.
\label{max-nn-concurrence}
\end{equation}     

According to the scaling ansatz~\cite{FINITE-SCALING} 
in the case of logarithmic 
singularities, the ratio between the two prefactors of the logarithm 
in Eqs.(\ref{nn-concurrence}) and (\ref{max-nn-concurrence}) is the exponent 
which governs the divergence 
of the correlation length $\xi$\cite{QUOTIENT}. 
In this case ($8/3\pi^2\approx 0.2702$) it follows that $\nu=1$, 
as it is known from the solution of the Ising model~\cite{PFEUTY}. 
By proper scaling~\cite{FINITE-SCALING}  and  taking into account the distance
of the minimum of $C(1)$ from the critical point, 
it is possible to make all the data for different $N$ collapse 
onto a single curve (Fig.\ref{fig2}). 
\puteps{fig4.eps}{8cm}{5cm}{
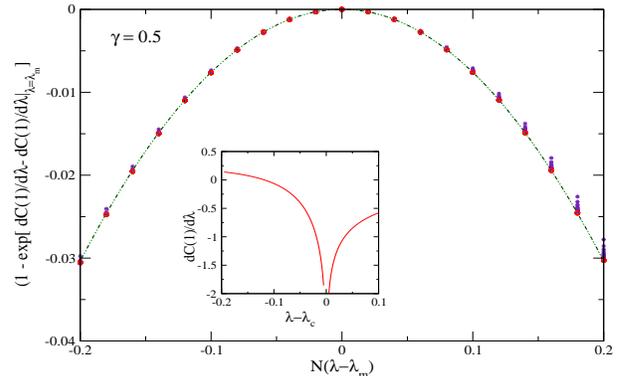
The universality hypothesis for the entanglement is checked
by considering the model Hamiltonian, defined in  Eq.(\protect\ref{model}), 
for a different value of $\gamma$. 
In this case we chose  $\gamma=0.5$ and $N$ ranging from $41$ up to $401$.
Data collapse, shown here for $C(1)$, is obtained for $\nu=1$, 
consistent with the model being in the universality class of the Ising model. 
In the inset is shown the divergence at the critical point 
for the infinite system.}{fig4}
This figure contains the data for the lattice size ranging from 
$N=41$ up to $N=2701$. 
These results show that all the key ingredients of the finite size scaling 
are present in the concurrence. We note that finite size 
scaling scaling is fulfilled over a very broad range of  values of $N$ which 
are of interest in several protocols in quantum information. 

A similar analysis can be carried on for the next-nearest neighbour 
concurrence $C(2)$. 
Since $\left. \partial_\lambda C(2)\right |_{\lambda_c}=0$, 
the logarithmic singularity here appears first in the second derivative 
with respect to $\lambda$ (see Fig.\ref{fig3} legend).
In the thermodynamic limit 
$ \partial_\lambda^2 C(2)=0.108\ln|\lambda-\lambda_c|+const$.
In this case also, the data collapse and finite size scaling (Fig.\ref{fig3})
agree with the expected scaling behaviour, $\nu=1$. 
This completes the analysis of entanglement for the one-dimensional 
Ising model.

A cornerstone in the theory of critical phenomena is the concept of 
universality -- that is, the critical properties depend only on the 
dimensionality of the system and the broken symmetry in the ordered phase. 
Universality in the critical properties of entanglement was verified 
considering the properties of the family of models defined in Eq.(1) 
with $\gamma=1/2$. The range of entanglement  $\xi_E$ is not universal. 
The maximum possible distance between entangled pairs increases and tends 
to infinity as $\gamma$ tends to zero. 
From the asymptotic behaviour of the reduced density matrix\cite{McCOY} 
we find that $\xi_E$ goes as $\gamma^{-1}$. This however has no 
dramatic consequences; the total concurrence $\sum_n C(n)$ stored in the 
chain is a weakly increasing function of $\gamma$ 
(for $0 < \gamma \le 1$, $0 < \sum_nC(n) < 0.2$). 
More interesting is the critical behaviour of the concurrence. 
To be specific we consider $C(1)$ in the case $\gamma = 0.5$, 
shown in Fig.4. As it was obtained for the Ising model,
scaling is fulfilled with the critical exponent $\nu=1$ in agreement 
with the universality hypothesis. 

The analysis of the 'resource' entanglement for a condensed matter system 
close to a quantum critical point allows to 
characterize both quantitatively and qualitatively the change
in the wave-function of the ground state on passing the phase transition. 
A notable feature which emerges is that though the entanglement itself 
is not an indicator of the phase transition, an intimate connection 
exists between entanglement, scaling and universality. In a way this
analysis allows us to discern what is genuinely quantum in a zero-temperature 
phase transition. 
The results presented here might be tested by
measuring different correlation functions, for example with neutron 
scattering, and extracting from these the entanglement properties of 
the ground state close to the critical point~\cite{THERMAL}. 

We finally discuss 
the perspectives of this work towards quantum computation. First,
system sizes considered here ($\sim 10^3$) could be those of a 
realistic quantum computer. Second, the scaling behaviour found could be
a powerful tool to evaluate (and hence to use) entanglement in
systems having different number of qubits.
In particular, close to the critical point, the entanglement depends strongly 
on the field -- so it could be tuned, realizing an 'entanglement switch'.
Last, long range correlations, typical of the critical region, might be 
of great importance in stabilizing the system against errors 
due to imperfections.

{\it Acknowledgments}. The authors would like to thank G.M. Palma, 
F. Plastina, and J. Siewert for helpful discussions.  
This work was supported by the European Community (IST-SQUIBIT), 
and by INFM-PRA-SSQI.

\end{multicols}

\end{document}